\documentclass[twoside]{dis04}
\def\runauthor{R. Peschanski}
\def\shorttitle{Saturation and Traveling Waves}
\newcommand{\rr}[4]{#1, {\it #2 \/}{\bf #3} #4}
\begin{document}

\title{SATURATION AND TRAVELING WAVES
}

\author{ROBI PESCHANSKI}

\address{Service de Physique Théorique, CEA/Saclay\\
URA 2306, unité de recherche associée au CNRS\\ 
 91191 Gif-sur-Yvette cedex, France\\
E-mail: pesch@spht.saclay.cea.fr}

\maketitle

\abstracts{High parton density effects with energy obey non-linear 
 QCD evolution equations for which exact solutions are not  known.  The 
mathematical class to which  the non-linear Balitsky-Kovchegov equation belongs is 
identified, proving the existence of asymptotic in energy traveling wave solutions 
which are ``universal'' i.e. independent of the initial conditions and of the 
precise form of the non-linearities. This has an direct impact on geometrical 
scaling and the diffusive transition to saturation, which is shown to be 
``normal'' for constant QCD coupling and ``abnormal'' for running coupling.}

\section{Introduction} 
Considering the scattering of a hard projectile (e.g. a massive QCD  dipole) on an 
extended  target, the Balitsky Fadin  Kuraev Lipatov (BFKL)  \cite{Lipatov:1976zz} 
evolution equation implies a densification of gluons and 
sea quarks with incident energy, while they keep in average the same size. It is thus 
natural to 
expect \cite {GLR} a modification of the evolution equation towards a {\it 
saturation} regime. Recently, a theoretical 
appoach  to saturation has been found \cite{venugopalan,balitsky} related  to  
non-linear  evolution equations of the gluon density in the framework of 
perturbative QCD. In the transition to saturation, the  exponential growth regime 
related to the  
BFKL kernel gets modified by   non-linear terms, leading to the Balitsky-Kovchegov 
(BK) equation \cite{balitsky}. A more general non-linear functional equation is 
expected to take into account the multiple correlations and to describe  the fully
saturated phase \cite{venugopalan}. The aim of our approach 
\cite{Munier1,Munier2,Munier3} 
is to explore the mathematical properties of the BK equation and derive its 
physical consequences for saturation in QCD.

\section{Saturation and Non-Linear Equations}

 The 
Balitsky-Kovchegov 
(BK) equation \cite{balitsky}  considers the energy evolution within the QCD 
dipole Hilbert space \cite{mueller1}.
To be specific let us consider
$N(Y,{x}_{01}),$  the dipole forward 
scattering amplitude and define
\begin{equation}
{\cal N}(Y,k)=\int_0^{\infty} \frac{dx_{01}}{x_{01}}
J_0(kx_{01})\,N(Y,x_{01})\ .
\label{eq:fourier}
\end{equation}

\begin{figure}[ht]
\centerline{\epsfxsize=3.9in\epsfbox{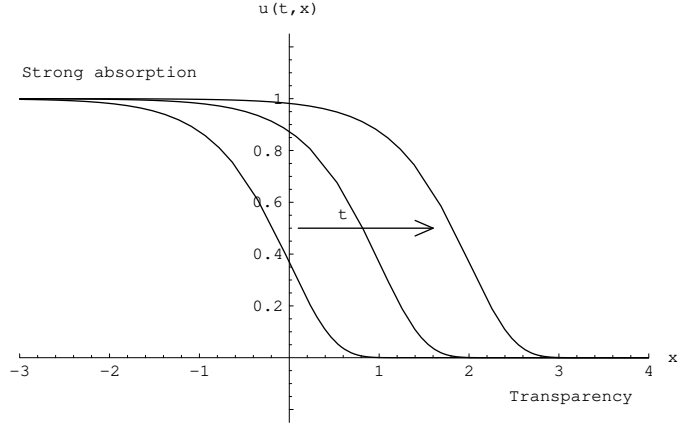}}   
\caption{{\it Typical traveling wave solution.} 
The function $u(t,x)$ is
represented for three different times. The wave front connecting the
regions
$u=1$ and $u=0$ travels
from the left
to the right as $t$ increases. That illustrates 
how the ``strong absorption'' or saturated phase
region invades the ``transparency'' region.
 \label{fig:fkpp}}
\end{figure}

Within suitable approximations
(large $N_c$, summation of fan diagrams, spatial homogeneity),  this quantity 
obeys (see the second reference in \cite{balitsky})
the nonlinear evolution equation
\begin{equation}
{\partial_Y}{\cal N}=\bar\alpha
\chi\left(-\partial_L\right){\cal N}
-\bar\alpha\, {\cal N}^2\ ,
\label{eq:kov}
\end{equation}
where $\bar\alpha=\alpha_s N_c/\pi$,
$\chi(\gamma)=2\psi(1)-\psi(\gamma)-\psi(1\!-\!\gamma)$ is the
characteristic function of the BFKL kernel \cite{Lipatov:1976zz},
$L=\log (k^2).$ 
In a first stage \cite{Munier1}, let us consider the  kernel  expanded up
to second order around $\gamma\!=\!{\scriptstyle \frac 12}.$

Eq.  (\ref{eq:kov}) boils down to a parabolic nonlinear
partial derivative equation:
\begin{equation}
{\partial_Y}{\cal N}=\bar\alpha
\left\{\chi\left({\scriptstyle \frac12}\right)+{\scriptstyle \frac12}
\chi^{\prime\prime} \left({\scriptstyle\frac12}\right)
\left(\partial_L+{\scriptstyle \frac12}\right)^2\right\}{\cal N}
-\bar\alpha\, {\cal N}^2\ .
\label{eq:kovchegov}
\end{equation}

The  mathematical point \cite{Munier1} of our recent approach is to 
remark that the structure 
of 
Eq.(\ref{eq:kovchegov}) is identical (by a suitable linear 
redefinition  ${\cal N}(L,Y) \to u(x,t)$  and for  fixed 
$\alpha$) to  the  Fisher and 
Kolmogorov-Petrovsky-Piscounov (F-KPP) equation \cite{KPP}:
\begin{equation}
\partial_t u(t,x)=\partial_x^2 u(t,x)+u(t,x)(1-u(t,x))
\label{eq:KPP}
\end{equation}
which appeared in the problem of gene diffusion and annihilation (1938). 

 \begin{figure}[ht]
\centerline{\epsfxsize=3.9in\epsfbox{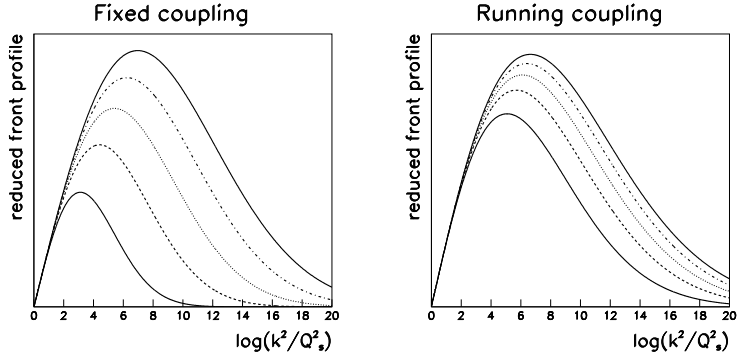}}   
\caption{{\it Evolution of the reduced front profile.} 
Fixed coupling: left; Running coupling: right.
The reduced front profile 
$(k^2/Q_s^2)^{\gamma_c}\,{\cal N}(k/Q_s(Y),Y)$ is plotted against
$\log(k^2/Q_s^2)$ for different rapidities.
 The various lines correspond to rapidities from 2 (lower
  curves, full line) up to 10 (upper curves).
Note the similarity of the wave fronts, but the quicker time evolution
(in $\sqrt{t}$) for fixed coupling, by contrast with the slow time
evolution (in $t^{1/3}$) for the running coupling case.
 \label{front}}
\end{figure}

\section{Universality and  Traveling Wave Solutions}
The  remarkable mathematical property of the 
F-KPP equation is  
 the existence of traveling wave solutions of the F-KPP 
equation \cite{KPP} at large times. This means that 
there exists a function 
of one variable $w$ such that 
\begin{equation}
\label{solution}
u(t{\scriptstyle{\rightarrow +\infty}},x)\  {\sim}\ 
w\left\{x-2 t-{\scriptstyle \frac32}\log t+{\cal O}(1)\right\}
\end{equation}
uniformly in $x$.
Such a solution is depicted on Fig.(\ref{fig:fkpp}).

This  analysis can be extended \cite{Munier2} to the study of the equation with the 
full kernel. Indeed, only the second-order expansion  around a given critical value 
$\gamma = {\gamma_c} = .6275...$ is relevant. Let us describe its general 
consequences. The well-known geometric scaling
property \cite{scaling} is obtained  for the solution of the 
non-linear equation 
(\ref{eq:kovchegov}) at large enough energy. 
In our notation, the geometric scaling property can be written
\begin{equation}
{\cal N}(Y,x_{01})={\cal N}\left(x_{01}{Q_s(Y)}\right)\ ,
\label{eq:defscaling}
\end{equation}
where
\begin{equation}
\label{Q}
 Q_s^2(Y) = 
\exp\left\{\bar\alpha\frac{\chi(\gamma_c)}{\gamma_c} Y -\frac{3}{2\gamma_c}\log Y
-\frac{3}{(\gamma_c)^2}
\sqrt{\frac{2\pi}{\bar\alpha\chi^{\prime\prime}(\gamma_c)}}\frac{1}{\sqrt{Y}} +  
{\cal O}({1/Y})
\right\}\ ,
 \end{equation}
  plays the role of  the saturation scale squared. Note that the solution
(\ref{solution}) mathematically requires  an  value initial condition ${\cal 
N}_0(k{\scriptstyle{\rightarrow +\infty}},Y_0) \ll 1/k^{2\gamma_c}$ which is realized 
in first order QCD by  color transparency ${\cal N}_0\sim 
1/k^2$. Note that the  result gives a rigorous proof of   previous  evaluations  based 
on linear evolution with boundary conditions (first term: \cite{iancu}, second term:  
\cite{Mueller:2002zm}); the third 
is new \cite{Munier3}.

It is possible to show that the result (\ref{solution}) is more general, by 
various extensions of the F-KPP solutions. First, some general arguments confirmed by 
numerical simulations (see the 
review in \cite{KPP}) lead to  expect the 
same result for the full nonlinear equation (\ref{eq:kov}). It is independent of 
the precise form of the non-linear damping terms and from the initial conditions 
(provided the transparency condition is fulfilled). Hence the ``Universality'' 
property. Second, 
the results can be extended to running $\alpha.$ One interesting  difference 
\cite{Munier2}
with the fixed $\alpha$ case  is the ``abnormal'' diffusion 
approach to scaling (in $t^{1/3}\sim Y^{1/6}$ instead of $t^{1/2}\sim Y^{1/2}$), see 
Fig.(\ref{front}).

\vspace{-.2cm}
\section*{Acknowledgements} The work and results described in this contribution 
were 
performed in collaboration with St\'ephane Munier (Centre de physique 
th{\'e}orique,
{\'E}cole polytechnique, 91128 Palaiseau cedex, France;
email: Stephane.Munier@cpht.polytechnique.fr).

\end{document}